\documentstyle[epsfig,12pt]{article}
\def\applt{\;\; {\lower3pt\hbox{$
{\buildrel < \over {\scriptstyle \sim} }$}}\;\;}

\large
\begin{document}

\centerline{\large \bf Transition to perturbative QCD
in two-photon collisions}

\vskip .4in

\centerline{Ron-Chou Hsieh$^a$\footnote{E-mail:
hsieh@hepth.phys.ncku.edu.tw} and Hsiang-nan Li$^{a,b}$\footnote{
E-mail: hnli@phys.sinica.edu.tw}} \vskip
0.4in

\centerline{${^a}$Department of Physics, National Cheng-Kung
University,} \centerline{Tainan, Taiwan 701, Republic of China}
\vskip 0.3cm

\centerline{$^b$Institute of Physics, Academia Sinica, Taipei,
Taiwan 115, Republic of China} \vbox {\vskip 1.0cm}


\vskip 1.0cm

\centerline{ABSTRACT} \vskip 0.5cm

We propose that the different angular distributions in two-photon
collisions observed at low and high center-of-mass energies
$W_{\gamma\gamma}$ indicate the transition from nonperturbative to
perturbative QCD. We calculate the differential cross sections of
$\gamma \gamma\to\pi \pi$, $KK$ in the angle $\theta$ of one of the
final-state mesons using QCD sum rules and the perturbative QCD
approach based on $k_T$ factorization theorem. Our predictions from
sum rules (perturbative QCD) decrease (increase) with $|\cos\theta|$,
consistent with the Belle data of $\gamma \gamma\to K^+K^-$ for
$W_{\gamma\gamma}\approx 1.5$-1.7 (2.2-2.4) GeV.


\vskip 1.0cm

It has remained as a controversy whether perturbative QCD (PQCD)
is applicable to exclusive processes at moderate energies
\cite{IL}. This issue has been widely discussed in the literature,
but general agreement is still not available. The discussion
usually focused on the simplest case, hadronic form
factors, which have been computed in nonperturbative
frameworks, such as QCD sum rules (QSR) \cite{SVZ}, and in the
PQCD formalism \cite{BL,ER,CZS,CZ}. Within theoretical
uncertainty, the predictions from both approaches were
claimed to be consistent with experimental data \cite{L98,SRJ}. The
contradictory conclusions on the dominant dynamics, soft
(Feynman mechanism \cite{NR}) or hard, in exclusive processes at
moderate energies were drawn. We have explained this subtlety
in \cite{KKLL}, and argued that there is no contradiction,
because the definitions of a soft contribution vary
among different theoretical frameworks. With this subtlety,
a hadronic form factor may not be the most appropriate quantity to
discriminate the soft-dominance and hard-dominance pictures.

In this work we shall propose that the angular distribution in
two-photon collisions is an appropriate quantity for the purpose.
We take the processes $\gamma\gamma\to \pi\pi$, $KK$ as an
example, and calculate the pion and kaon angular distributions
using nonperturbative QSR, which are reliable at a low
center-of-mass energy $W_{\gamma\gamma}$, and the PQCD approach
based on $k_T$ factorization theorem \cite{BS,LS,NL,HS,RP}, which
is reliable at a high $W_{\gamma\gamma}$. It will be shown that
QSR and PQCD give dramatically different predictions: the former
decrease, while the latter increase with $|\cos\theta|$, where
$\theta$ denotes the angle of one of the final-state mesons in the
center-of-mass frame. In fact, this difference is a general
feature of QSR and PQCD predictions for two-photon collisions,
regardless the final states being baryons or mesons. It is thus
important to measure the dependence of the angular distribution on
$W_{\gamma\gamma}$, and to see whether it evolves with
$W_{\gamma\gamma}$ following the QSR and PQCD predictions. Such a
transition has indeed been observed in the process
$\gamma\gamma\to K^+K^-$ by Belle recently \cite{BelKK}.

It will be shown that our predictions from QSR (PQCD) are
consistent with the experimental data of the $\gamma\gamma\to
K^+K^-$ differential cross section \cite{BelKK} for
$W_{\gamma\gamma}\approx 1.5$-1.7 (2.2-2.4) GeV. Hence, we
conclude that the transition from nonperturbative to perturbative
QCD in $\gamma\gamma\to K^+K^-$ occurs at $W_{\gamma\gamma}\approx
2$ GeV, the same as that drawn from the analysis of the pion form
factor \cite{LS}. The transition was not observed by ALEPH,
since only the $\gamma\gamma\to \pi^+\pi^-$, $K^+K^-$ cross sections
with $W_{\gamma\gamma}> 2$ GeV were measured \cite{ALEPH}.
The PQCD calculation of the processes
$\gamma\gamma\to \pi\pi$, $KK$ has been performed in \cite{CV},
but smaller cross sections were obtained. We shall point out that
the difference between this work and \cite{CV} comes from the
models of meson distribution amplitudes. We shall also compare the
PQCD approach with another different type of factorization theorem
formulated by means of two-meson distribution amplitudes
\cite{DKV}, both of which give similar predictions.
The same transition has been also observed in
$\gamma \gamma\to p \bar p$ \cite{KUO}: the proton angular
distribution decreases with $|\cos\theta|$ for $W_{\gamma\gamma}<
2.5$ GeV and becomes increasing for $W_{\gamma\gamma}> 2.5$ GeV.
The latter behavior is consistent with the predictions from the
PQCD approach based on $k_T$ factorization theorem \cite{H} and
from the diquark model \cite{Kro93}, in which a hard scattering
takes place between the diquark and the third valence quark. Note
that the transition scale was claimed to be about 3 GeV from the
analysis of the proton form factor \cite{HN}.


The differential cross section of pion Compton scattering
$\pi\gamma\to\pi\gamma$ has been calculated in QSR and in PQCD
\cite{CRS,co,CL93,CL95}. The $\gamma\gamma\to \pi\pi$ differential
cross section can be easily derived by exchanging the
Mandelstam invariants $s$ and $t$. Since the QSR formulas in
\cite{CL95} contain incomplete sub-leading terms in powers of $1/s$,
$1/t$, the straightforward exchange of $s$ and $t$ does not
respect the symmetry under the reflection between $\cos\theta$ and
$-\cos\theta$. One must neglect the sub-leading terms first, and
then exchange $s$ and $t$. The resultant differential
cross section for two-photon collision $\gamma(q_1)\gamma(q_2)\to
\pi(p_1)\pi(p_2)$ in QSR is written as,
\begin{equation}
\frac{d \sigma}{d\cos\theta}=\frac{|{\cal M}|^2}{32\pi
s}\;,\;\;\;\; |{\cal M}|^2=\frac{|H_1|^2+|H_2|^2}{2}\;,
\label{csc}
\end{equation}
where the amplitudes $H_i$, $i=1$, 2, come from the decomposition
of the amplitude,
\begin{eqnarray}
{\cal M}_{\mu\nu}(s,t,u)= H_1(s,t,u) e^{(1)}_{\mu}e^{(1)}_{\nu} +
H_2(s,t,u) e^{(2)}_{\mu}e^{(2)}_{\nu}\;,
\label{h1h2}
\end{eqnarray}
with the subscripts $\mu$ and $\nu$ corresponding to the two photn
vertices. The helicity vectors $e^{(1)}$ and $e^{(2)}$ have been
defined in \cite{CRS,CL93}, which satisfy the orthogonality conditions
$e^{(i)}\cdot e^{(j)}=-\delta_{ij}$.

The explicit expressions of $H_i$ are given by
\begin{eqnarray}
&&{f_\pi}^2H_i(s,t,u){tu\over s}=
\left[\int_{0}^{s_0}ds_1\int_{0}^{s_0}ds_2
{\rho_i}^{\rm pert}(s,t,u,s_1,s_2)\right.
\nonumber \\
& &\left.\hspace*{0.5cm}
+\frac{\alpha_s}{\pi}\langle G^2 \rangle
\int_{0}^{s_0}ds_1\int_{0}^{s_0}ds_2
\rho_i^{\rm gluon}(s,t,u,s_1,s_2)\right]
\exp[-(s_1+s_2)/M^2] \nonumber \\
& &\hspace*{0.5cm}+C_i^{\rm quark}(s,t,u)\pi\alpha_s \langle
(\bar{\psi}\psi)^2 \rangle\;, \label{h2}
\end{eqnarray}
$f_\pi=132$ MeV being the pion decay constant.
The Mandelstam invariants are given by
\begin{eqnarray}
& &s=(q_1+q_2)^2=W_{\gamma\gamma}^2\;,
\nonumber\\
& &t=(q_1-p_1)^2=-\frac{W_{\gamma\gamma}^2}{2}(1-\cos\theta)\;,
\nonumber\\
& &u=(q_1-p_2)^2=-\frac{W_{\gamma\gamma}^2}{2}(1+\cos\theta)\;.
\end{eqnarray}
For the process $\gamma(q_1)\gamma(q_2)\to K(p_1)K(p_2)$, the
differential cross section is obtained from Eq.~(\ref{h2}) by
applying the replacements,
\begin{eqnarray}
f_\pi \to f_K\;,\;\;\;\; \exp[-(s_1+s_2)/M^2]\to
\exp[(2m_K^2-s_1-s_2)/M^2]\;,
\end{eqnarray}
with the kaon decay constant $f_K=160$ MeV and the kaon mass
$m_K=0.49$ GeV, which is not negligible compared to the value of
$s_0$ determined below.

The method to evaluate the perturbative spectral densities
${\rho_i}^{\rm pert}$ and the power corrections, $\rho_i^{\rm gluon}$
and $C_i^{\rm quark}$, has been described in \cite{co}.
The perturbative, gluonic and quark contributions to $H_1$ are
written, respectively, as
\begin{eqnarray}
\rho_1^{\rm pert}&=&{640Q^{14}(Q^{2}-t)(Q^2-u)(s_1+s_2)\over
3\pi^2 (4Q^4-s_1s_2)^5}\;,
\nonumber\\
\rho_1^{\rm gluon}&=&{5120 Q^{20}(4tu + 9 Q^4) \over 27
(2Q^2-s_1)^2(2Q^2-s_2)^2(4Q^4-s_1s_2)^5}\;,
\nonumber\\
C_1^{\rm quark} &=&-{16\over 9} {(t-u)^2\over M^4 s}\;,
\label{h1p}
\end{eqnarray}
for the variable,
\begin{eqnarray}
Q^2=\frac{1}{4}\left(s_1+s_2-s+\sqrt{(s_1+s_2-s)^2-4s_1s_2}\right)\;.
\label{qv}
\end{eqnarray}
The corresponding quantities associated with $H_2$ are given by
\begin{eqnarray}
\rho_2^{\rm pert}&=&{-640 Q^{14}(Q^{2}-t)(Q^2-u)(s_1+s_2)\over
\pi^2 (4Q^4-s_1s_2)^5}\;,
\nonumber\\
\rho_2^{\rm gluon}&=&{-5120Q^{20}(12tu + 7 Q^4) \over 27
(2Q^2-s_1)^2(2Q^2-s_2)^2(4Q^4-s_1s_2)^5}\;,
\nonumber \\
C_2^{\rm quark}&=&-\frac{128}{9}\frac{tu}{M^2s^2}\;. \label{h2p}
\end{eqnarray}
It is easy to find that Eqs.~(\ref{h1p}) and (\ref{h2p})
respect the symmetry under the exchange of $t$ and $u$, i.e., of
$\cos\theta$ and $-\cos\theta$. The gluon and quark condensates,
$\langle G^2\rangle$ and $\langle (\bar{\psi}\psi)^2\rangle$, take
the values
\begin{eqnarray}
&&\frac{\alpha_s}{\pi}\langle G^2\rangle=
1.2\times 10^{-2}{\rm GeV}^4\;,\nonumber \\
&&\alpha_s\langle (\bar{\psi}\psi)^2\rangle
=1.8\times 10^{-4}{\rm GeV}^6\;,
\label{vev}
\end{eqnarray}
respectively.

In the PQCD approach based on $k_T$ factorization theorem, we have
derived the factorization formulas for pion Compton scattering
\cite{CL95}. Compared to the derivation in collinear factorization
theorem \cite{ZM,KN,FSZ}, parton transverse momenta have been retained.
Note that there are minor mistakes in the factorization formulas for
the pion Compton scattering presented in \cite{CL95}. We have taken
this chance to correct them, and then applied the exchange of $s$ and
$t$. The resultant expressions for $H_i$, $i=1$ and 2, involved in the
two-photon collision $\gamma\gamma\to \pi\pi$ are given by
\begin{eqnarray}
H_i(s,t,u)&=&\int_{0}^{1}d x_{1}d x_{2}\,
\phi_\pi(x_{1})\phi_\pi(x_{2}) \int_{0}^{\infty} bd b
\left[(e_u^2+e_d^2)T_{i}(x_i,s,t,u,b)\right.\nonumber \\
& &\left.+2e_ue_dT_{i}^{\prime}(x_i,s,t,u,b)\right]
\exp[-S(x_i,b,W_{\gamma\gamma}/\sqrt{2})]\; ,
\label{15}
\end{eqnarray}
where the variable $b$ denotes the transverse separation
between the two valence quarks of the pion, and the charge factors
are $e_u^2+e_d^2=5/9$ and $e_ue_d=2/9$. The corresponding hard kernels
are written as
\begin{eqnarray}
T_1&=&-\frac{32\pi^2
C_F\alpha\alpha_s(w_1)}{(1-x_1)(1-x_2)}
\left(\frac{u}{t}+\frac{t}{u}+4-2x_1-2x_2\right)
\frac{i\pi}{2}H_0^{(1)}\left(\sqrt{r_1}b\right)\;,
\nonumber\\
T'_1&=&32\pi^2C_F\alpha\alpha_s(w_2)
\left[\theta(-r_2)K_0\left(\sqrt{|r_2|}b\right)
+\theta(r_2)\frac{i\pi}{2}H_0^{(1)}\left(\sqrt{r_2}b\right)\right]
\nonumber\\
& &\times\left[\frac{1}{x_1(1-x_1)}+\frac{1}{x_2(1-x_2)}
+\frac{s}{x_2(1-x_1)t}+\frac{s}{x_1(1-x_2)u}\right]\;,
\label{1hh}\\
T_2&=&T-T_1\;,\;\;\;\;T'_2=T'-T'_1\;,
\nonumber\\
T &=&\frac{64\pi^2
C_F\alpha\alpha_s(w_1)}{(1-x_1)(1-x_2)}
\frac{i\pi}{2}H_0^{(1)}\left(\sqrt{r_1}b\right)
\left\{\frac{[(1-x_1)s+u][(1-x_2)s+u]}{t^2} \right.
\nonumber \\
& &\left.+\frac{[(1-x_1)s+t][(1-x_2)s+t]}{u^2}
-2(1-x_1)-2(1-x_2)\right\}\;,
\nonumber\\
T'&=&64\pi^2C_F\alpha\alpha_s(w_2)\left[\theta(-r_2)
K_0\left(\sqrt{|r_2|}b\right)
+\theta(r_2)\frac{i\pi}{2}H_0^{(1)}\left(\sqrt{r_2}b\right)\right]
\nonumber \\
&  & \times\left[\frac{1}{x_1(1-x_1)}-
\frac{(1+x_2-x_1 x_2)s^2+(1+x_2-x_1)su}{x_2(1-x_{1})t^2}\right.
\nonumber \\
& &\;\;\;\;\left.+\frac{1}{x_2(1-x_2)}-
\frac{(1+x_1-x_1x_2)s^2+(1+x_1-x_2)st}{x_1(1-x_{2})u^2}\right]\;,
\label{hh}
\end{eqnarray}
with the constant $\alpha=1/137$, the color factor $C_F=4/3$,
the Bessel functions $K_{0}$ and $H_0^{(1)}$, and the invariants,
\begin{eqnarray}
r_1=x_1x_2s,\;\;\;\;\;\;r_2=x_{1}x_{2}s+x_1u+x_2t\; .
\end{eqnarray}
The arguments $w_l$ of $\alpha_s$ are chosen as the largest mass
scales in the hard scattering,
\begin{equation}
w_1=\max\left(\sqrt{r_1},\frac{1}{b}\right),\;\;\;\;\;\;
w_2=\max\left(\sqrt{r_2},\frac{1}{b}\right)\; . \label{9}
\end{equation}
The extra Sudakov factor $\exp(-S)$ compared to the standard
collinear factorization formula \cite{BL} arises from the
all-order summation of the large logarithms in $k_T$ factorization
theorem \cite{LL04}. Simply speaking, it describes the extrinsic
$b$ dependence of a parton in the pion, and decreases quickly in
the large-$b$ region. This is how the Sudakov suppression improves
the perturbative expansion. There also exists an intrinsic $b$
dependence \cite{Kro}, which is less essential compared to the
Sudakov effect in the processes involving only
light mesons. For the explicit expression of $\exp(-S)$
and the values of the QCD scale $\Lambda_{\rm QCD}$ ($=0.25$ GeV)
and of the quark flavor number $n_f$ ($=3$), refer to \cite{KLS}.


The PQCD factorization formulas for $\gamma\gamma\to K^+K^-$
are slightly different. Since the kaon distribution amplitude
is not symmetric under the exchange of $x$ and $1-x$, the contributions
from the diagrams with the photons attaching the $u$ quark and the $s$
quark can not be combined. The amplitudes are expressed as
\begin{eqnarray}
H_i(s,t,u)&=&\int_{0}^{1}d x_{1}d x_{2}\,\int_{0}^{\infty} bd b
\left\{[e_s^2T_{i}(x_i,s,t,u,b)
+e_ue_sT_{i}^{\prime}(x_i,s,t,u,b)]\phi_K(x_{1})\phi_K(x_{2})
\right.\nonumber\\
& &\left.+[e_u^2T_{i}(x_i,s,t,u,b)
+e_ue_sT_{i}^{\prime}(x_i,s,t,u,b)]\phi_K(1-x_{1})\phi_K(1-x_{2})
\right\}
\nonumber \\
& & \times \exp[-S(x_i,b,W_{\gamma\gamma}/\sqrt{2})]\; ,
\label{16}
\end{eqnarray}
with $\phi_K$ being the kaon distribution amplitude.
The pion and kaon distribution amplitudes are adopted as \cite{PB1,BB},
\begin{eqnarray}
\phi_\pi(x)&=&\frac{3f_{\pi}}{\sqrt{2N_{c}}}\,x(1-x)
\left[1+0.44C_2^{3/2}(1-2x)+0.25C_4^{3/2}(1-2x)\right]\; ,
\label{asb}\\
\phi_K(x)&=&\frac{3f_{K}}{\sqrt{2N_{c}}}\,x(1-x)
[1-0.54(1-2x)+0.16C_2^{3/2}(1-2x)]\; , \label{ck}
\end{eqnarray}
with $N_{c}=3$ being the number of colors, and the Gegenbauer
polynomials,
\begin{equation}
C_2^{3/2}(t)=\frac{3}{2}(5t^2-1)\;,\;\;\;
C_4^{3/2}(t)=\frac{15}{8}(21 t^4 -14 t^2 +1) \;.
\end{equation}


We then perform the numerical analysis of the pion and
kaon angular distributions for various center-of-mass energy
$W_{\gamma\gamma}$ in two-photon collisions. For QSR,
the duality interval $s_0$ is determined in the way that
the amplitudes $H$ are most stable with respect to the variation
of the Borel mass $M^2$. The best choice of $s_0$ are found to be
$s_0=0.54$, 0.57, 0.59, 0.615, and 0.675 GeV$^2$ at
$W_{\gamma\gamma}=1.50$, 1.54, 1.58, 1.62, and 1.70 GeV,
respectively, for which the QSR results become approximately constant
as $M^2>2$ GeV$^2$. Hence, we have chosen $M^2=4$ GeV$^2$ in
Eq.~(\ref{h2}). Since $s_0$ increases with $W_{\gamma\gamma}$,
higher excitations beyond the pion pole will contribute to
the left-hand sides of Eq.~(\ref{h2}) at some large value of
$W_{\gamma\gamma}$, such that the sum rules fail. Therefore,
we regard the QSR results for $W_{\gamma\gamma}<1.7$ GeV as being
reliable. Note that the best choice of $s_0$ slightly depends on
$|\cos\theta|$ for a fixed $W_{\gamma\gamma}$, varying in the
range of $\pm 0.005$ GeV$^2$.

The QSR curves for the $\gamma\gamma\to\pi\pi$, $KK$ differential
cross section $d \sigma/d\cos\theta$ are shown in Fig.~1, which
exhibit a decrease with $|\cos\theta|$, consistent with the
tendency of the Belle data \cite{BelKK}. Because the pion decay
constant is smaller than the kaon one, the $\gamma\gamma\to\pi\pi$
cross sections are larger according to Eq.~(\ref{h2}). We have
noticed that the QSR results are very sensitive to the variation
of $s_0$, since, as indicated in Eqs.~(\ref{h1p}) and (\ref{h2p}),
the perturbative spectral densities behave like $s_0^{-2}$ for
$Q^2\sim s_1\sim s_2\sim s_0$. $Q^2$ is small in two-photon
collisions, but of order $t$ in Compton scattering, which can be
realized from Eq.~(\ref{qv}) by substituting the negative $t$ for
$s$. This is the reason the QSR results for the latter are
insensitive to $s_0$ \cite{CL95}. To demonstrate the sensitivity,
we present three curves in Fig.~1, corresponding to the best $s_0$
plus 0.005 GeV$^2$, to the best $s_0$, and to the best $s_0$ minus
0.005 GeV$^2$ from top to bottom, respectively, which survive the
stability analysis for different $\theta$ as mentioned above. The
range enclosed by these three curves represents the theoretical
uncertainty, within which the QSR predictions are in agreement
with the $\gamma\gamma\to K^+K^-$ data \cite{BelKK}. We stress
that as long as a stability window exists, QSR results should be
regarded as being reliable. Therefore, the results presented here
are solid, though they exhibit larger theoretical uncertainty. It
is found that the deviation from the data becomes, as expected,
more obvious as $W_{\gamma\gamma}>1.7$ GeV.

\begin{figure}
\begin{center}
\includegraphics[height=20cm]{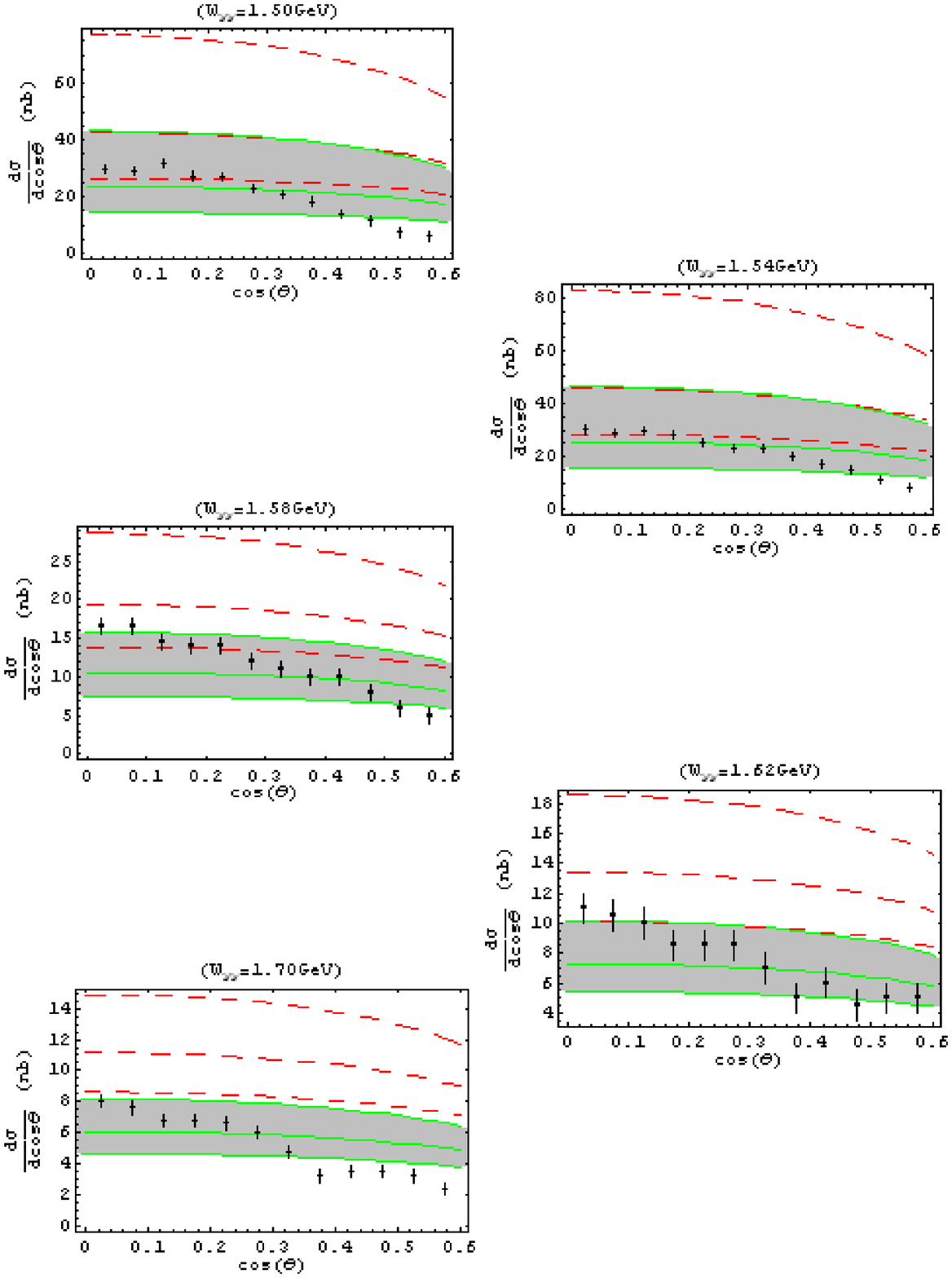}
\end{center}
\caption{\label{fig1}%
$d \sigma/d\cos\theta$ derived from QCD sum rules for
$W_{\gamma\gamma}=1.50$, 1.54, 1.58, 1.62, and 1.70 GeV with the
dashed (solid) lines corresponding to $\gamma\gamma\to \pi^+\pi^-$
($\gamma\gamma\to K^+K^-$). The three curves in each
plot correspond to the best $s_0$ plus 0.005 GeV$^2$, to the best
$s_0$, and to the best $s_0$ minus 0.005 GeV$^2$ from top to bottom,
respectively. The area enclosed by the three curves for
$\gamma\gamma\to K^+K^-$ has been shaded. The data points arise from
the range of $W_{\gamma\gamma}\pm 0.02$ GeV.}
\end{figure}

The $\gamma\gamma\to\pi\pi$, $KK$ differential cross section
derived from PQCD for $W_{\gamma\gamma}=2.22$, 2.26, 2.30, 2.34,
and 2.38 GeV are shown in Fig.~2, which exhibit an increase with
$|\cos\theta|$, also consistent with the tendency of the Belle
data \cite{BelKK}. The ascending of the PQCD results is understood
through the hard kernels, which are proportional to $1/(tu)\propto
1/(1-\cos^2\theta)$. The $\gamma\gamma\to KK$ cross sections are
smaller despite of the decay constants $f_K>f_\pi$, since the
amplitudes with the photons attaching the $s$ quark are suppressed
by the shape of the kaon distribution amplitude. It is found that
the PQCD predictions are in good agreement with the
$\gamma\gamma\to K^+K^-$ data \cite{BelKK}, and that the deviation
from the data becomes more obvious at lower $W_{\gamma\gamma}$:
Figure 2 shows that the data descend a bit first before ascending
with $|\cos\theta|$ for $W_{\gamma\gamma}=2.22$ GeV, a behavior
whose explanation requires a theoretical framework more
sophisticated than QSR and PQCD. We conclude that the high-energy
behavior of two-photon collisions can be explained perfectly by
PQCD.

Note that our results are larger than those obtained from a
similar PQCD analysis based on $k_T$ factorization theorem in
\cite{CV}, due to the different models of meson distribution
amplitudes. If employing the asymptotic model of the pion
distribution amplitude favored by \cite{CV,KR},
\begin{eqnarray}
\phi_\pi^{AS}(x)&=&\frac{3f_{\pi}}{\sqrt{2N_{c}}}\,x(1-x)\;,
\label{as}
\end{eqnarray}
our numerical results shown in Fig.~2 will be close to those in
\cite{CV}. We would like to mention that the above asymptotic
model has been excluded by \cite{BMS}. The completely
contradictory conclusions are attributed to the different
treatments of the subleading contributions, when the pion
distribution amplitude was extracted from the data of the pion
transition form factor. The authors in \cite{KR} considered only
the leading-order hard kernel and the leading-twist (twist-2) pion
distribution amplitude, and included the Sudakov factor. However,
those in \cite{BMS} further considered the next-to-leading-order
hard kernel and the twist-4 pion distribution amplitude, but did
not take into account the Sudakov factor (due to the fast
decreasing behavior of their distribution amplitudes at the end
points of the momentum fraction). For other postulations for the
models of the pion distribution amplitude, refer to \cite{SSK}.

The Gegenbauer coefficients of the kaon distribution amplitude in
Eq.~(\ref{ck}) also exhibit uncertainty from the sum-rule analysis
\cite{BB}. We have examined that our predictions are insensitive
(within 10\%) to the variation of the first coefficient between
$0.54\pm 0.27$, but are to the variation of the second coefficient
between $0.16\pm 0.10$. Therefore, we display three solid curves
for the $\gamma\gamma\to KK$ differential cross section
corresponding to the second coefficient 0.26, 0.16 and 0.06 from
top to bottom, respectively. The area enclosed by these three
curves can be regarded as part of the theoretical uncertainty in
the PQCD calculation. Strictly speaking, a Gegenbauer coefficient
evolves with the scale $1/b$ in $k_T$ factorization theorem
governed by $[\alpha_s(1/b)/\alpha_s(\mu_0)]^\gamma$, where
$\mu_0=1$ GeV represents the initial scale the evolution starts
with, and $\gamma$ is an anomalous dimension \cite{BB}. We have
investigated this evolution effect, and found that it is not
essential and can be covered by the theoretical uncertainty from
the variation of the Gegenbauer coefficient.

\begin{figure}
\begin{center}
\includegraphics[height=20cm]{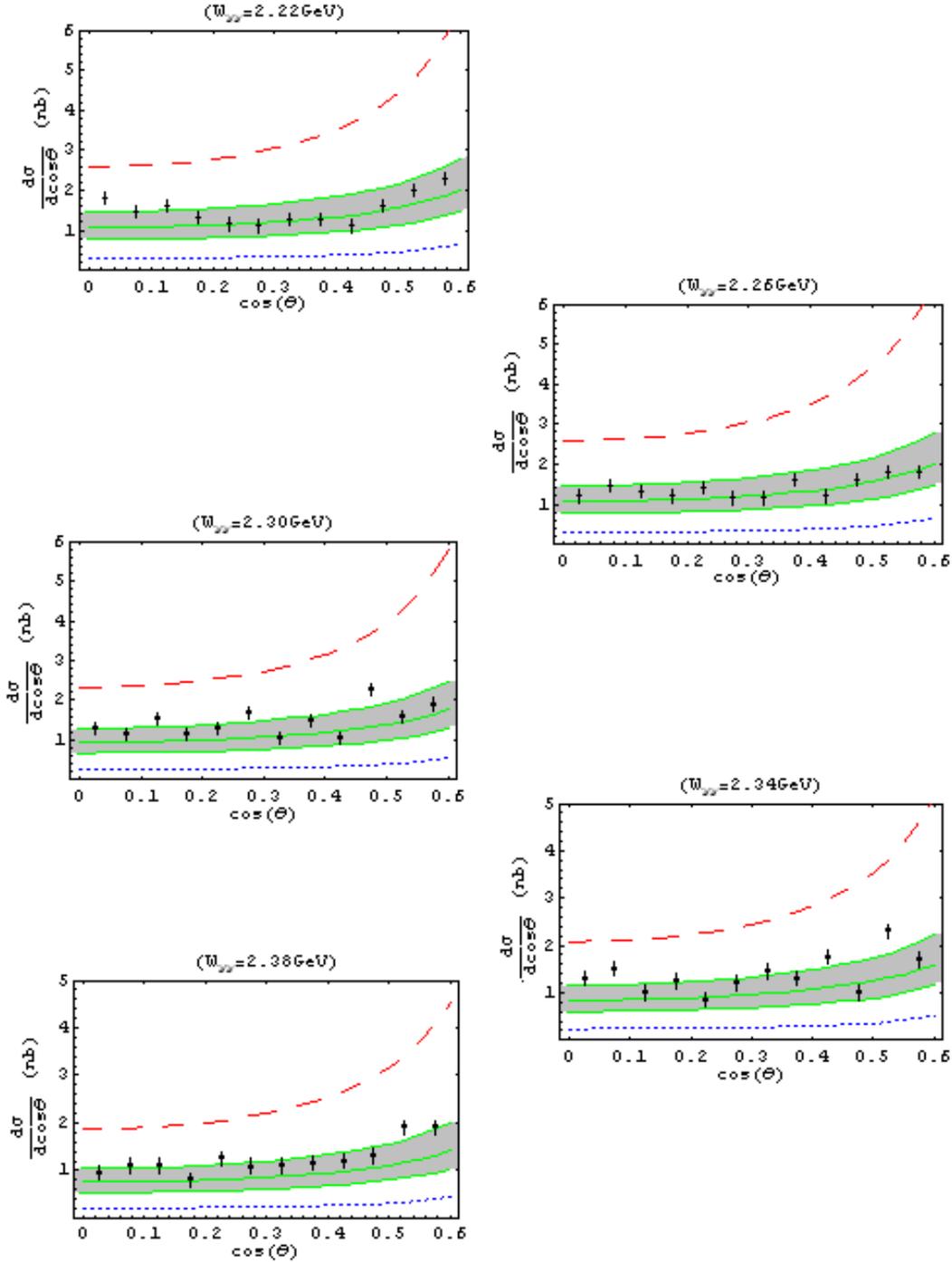}
\end{center}
\caption{\label{fig2}%
$d \sigma/d\cos\theta$ derived from PQCD based on $k_T$
factorization theorem for $W_{\gamma\gamma}=2.22$, 2.26, 2.30,
2.34, and 2.38 GeV. The dashed (dotted) lines for $\gamma\gamma\to
\pi^+\pi^-$ correspond to the pion distribution amplitude in
Eq.~(\ref{asb}) [Eq.~(\ref{as})]. The three solid curves for
$\gamma\gamma\to K^+K^-$ correspond to the second Gegenbauer
coefficients 0.26, 0.16 and 0.06 of the kaon distribution
amplitude from top to bottom, respectively.}
\end{figure}

Below we compare the PQCD approach with another different type of
factorization theorem formulated by means of two-meson
distribution amplitudes \cite{DKV}. In the latter formalism the
hard kernel is represented by the $\gamma\gamma\to q\bar q$
scattering, and the nonperturbative input is a two-meson
distribution amplitude, which collects soft gluon exchanges
between the energetic $q\bar q$ pair. Since this factorization
theorem involves the hard scattering represented by quark
diagrams, its prediction for the angular distributions should also
show the characteristic of the spin-half particle production,
i.e., increase with $|\cos\theta|$ as in PQCD. It has been claimed
\cite{DKV} that the resultant bag-diagram contribution dominates
over that from PQCD. First, we point out that the bag-diagram
contribution is sufficient to account for the
$\gamma\gamma\to\pi\pi$, $KK$ cross sections, because the
two-meson distribution amplitudes have been tuned to fit the data.
The PQCD contribution is small, because the asymptotic model of
the meson distribution amplitudes was adopted in \cite{CV}.
Therefore, the dominance of the bag-diagram contribution requires
an independent check, that is, an independent study of the
behavior of the two-meson distribution amplitudes using
nonperturbative methods. Second, there is an overlap between the
bag-diagram and PQCD predictions, and a comparison between them
should be made carefully. The quark diagrams in PQCD, in which the
two photons attach the same quark lines and the exchanged gluon
attaches the two valence quarks of one of the mesons, can be
factorized following \cite{DKV}, when the spectator quarks carry
small momentum fractions. It is not necessary to specify a small
momentum fraction, at which the bag-diagram factorization holds,
since it is related to a factorization scheme, and quite
arbitrary. We have investigated the contribution from these
diagrams and found that it exceeds half of the differential cross
section only in the large $|\cos\theta|$ ($> 0.5$) region. Anyway,
an experimental discrimination of the bag-diagram and PQCD
approaches has been proposed in \cite{DKV}: the equality of the
$\gamma\gamma\to\pi^+\pi^-$ and $\gamma\gamma\to\pi^0\pi^0$ cross
sections would favor the former.


In this paper we have calculated the pion and kaon angular
distributions $d\sigma/d\cos\theta$ in the two-photon collisions
$\gamma\gamma\to \pi\pi$, $KK$ using both the nonperturbative QSR
and the PQCD approach based on $k_T$ factorization theorem. It has
been shown that the predicted angular distributions in the two
approaches differ dramatically: the former decrease, while the
latter increase with $|\cos\theta|$. Therefore, the change from
the descending behavior at a low center-of-mass energy
$W_{\gamma\gamma}$ to the ascending behavior at a high
$W_{\gamma\gamma}$ indicates the transition from nonperturbative
dynamics to perturbative dynamics. An angular distribution in
two-photon collisions thus serves the purpose for discriminating
the soft-dominance and hard-dominance pictures, which is more
unambiguous than a hadronic form factor usually considered in the
literature. Such a transition has indeed been observed by Belle at
$W_{\gamma\gamma}\approx$ 2 GeV in $\gamma\gamma\to K^+K^-$
\cite{BelKK}, and at $W_{\gamma\gamma}\approx$ 2.5 GeV in
$\gamma\gamma\to p\bar p$ \cite{KUO}. Our results from QSR (PQCD)
are consistent with the $\gamma\gamma\to K^+K^-$ data for
$W_{\gamma\gamma}\approx 1.5$-1.7 (2.2-2.4) GeV.

\vskip 1.0cm
We thank A. Chen, C.C. Kuo, A. Finch, P. Kroll, and
N.G. Stefanis for useful discussions. This work
was supported by the National Science Council of R.O.C. under Grant
No. NSC-92-2112-M-001-030.


\begin{thebibliography}{99}

\bibitem{IL} N. Isgur and C.H. Llewellyn Smith, Nucl. Phys.
{\bf B317}, 526 (1989).
\bibitem{SVZ} M.A. Shifman, A. I. Vainshtein and V.I. Zakharov, Nucl.
Phys. {\bf B147}, 385, 448, 519 (1979).
\bibitem{BL} G.P. Lepage and S.J. Brodsky, Phys. Lett. B {\bf 87},
359 (1979); Phys. Rev. D {\bf 22}, 2157 (1980).
\bibitem{ER} A.V. Efremov and A.V. Radyushkin, Phys. Lett. B {\bf 94},
245 (1980).
\bibitem{CZS} V.L. Chernyak, A.R. Zhitnitsky, and V.G. Serbo,
JETP Lett. {\bf 26}, 594 (1977).
\bibitem{CZ} V.L. Chernyak and A.R. Zhitnitsky,
Sov. J. Nucl. Phys. {\bf 31}, 544 (1980); Phys. Rep. {\bf 112},
173 (1984).
\bibitem{L98} B. Kundu {\it et al.}, Eur. Phys. J. C {\bf 8},
637 (1999); P. Jain {\it et al.}, Nucl. Phys. {\bf A666}, 75 (2000);
H-n. Li, Nucl. Phys. {\bf A684}, 304 (2001).
\bibitem{SRJ} A. Szczepaniak, A. Radyushkin, and C.R. Ji,
Phys. Rev. D {\bf 57}, 2813 (1998).
\bibitem{NR}V.A. Nesterenko and A.V. Radyushkin, Phys. Lett. B {\bf
115}, 410 (1982); A.V. Radyushkin, Acta Phys. Polon. B {\bf 15},
403 (1984).
\bibitem{KKLL} Y.Y. Keum, T. Kurimoto, H-n. Li, C.D. Lu, and A.I.
Sanda, hep-ph/0305335, to appear in Phys. Rev. D.
\bibitem{BS} J. Botts and G. Sterman, Nucl. Phys. {\bf B325}, 62
(1989).
\bibitem{LS} H.N. Li and G. Sterman, Nucl. Phys. {\bf B381},
129 (1992); D. Tung and H-n. Li, Chin. J. Phys. {\bf 35}, 651 (1997).
\bibitem{NL} H-n. Li, Phys. Rev. D {\bf 55}, 105 (1997);
D {\bf 64}, 014019 (2001); hep-ph/0304217;
M. Nagashima and H-n. Li, hep-ph/0202127; Phys.
Rev. D {\bf 67}, 034001 (2003).
\bibitem{HS} T. Huang and Q.X. Shen, Z. Phys. C {\bf 50}, 139 (1991).
\bibitem{RP} J.P. Ralston and B. Pire, Phys. Rev. Lett. {\bf 65},
2343 (1990).
\bibitem{BelKK} Belle Collaboration, K. Abe {\it et al.}, Eur.
Phys. J. C {\bf 32}, 323 (2004).
\bibitem{ALEPH} ALEPH Collaboration, A. Heister {\it et al.},
Nucl. Phys. {\bf B569}, 140 (2003).
\bibitem{CV} C. Vogt, hep-ph/0010040.
\bibitem{DKV} M. Diehl, P. Kroll, and C. Vogt, Phys. Lett. B
{\bf 532}, 99 (2002); Eur. Phys. J. C {\bf 26}, 567 (2003).
\bibitem{KUO} C.C. Kuo, Belle Note No. 691.
\bibitem{H} T. Hyer, Phys. Rev D {\bf 47}, 3875 (1993).
\bibitem{Kro93} P. Kroll, Th. Pilsner, M. Schuermann, and W.
Schweiger, Phys. Lett. B {\bf 316}, 546 (1993).
\bibitem{HN} H.N. Li, Phys. Rev. D {\bf 48}, 4243 (1993); B. Kundu,
H-n. Li, J. Samuelsson, and P. Jain, Eur. Phys. J. C {\bf 8}, 637 (1999).
\bibitem{CRS} C. Corian\`{o}, A. Radyushkin and G. Sterman,
 Nucl. Phys. {\bf B405}, 481 (1993).
\bibitem{co} C. Corian\`{o}, Nucl. Phys. {\bf B410}, 90 (1993).
\bibitem{CL93} C. Corian\`{o} and H-n. Li, Phys. Lett. B {\bf 309},
409 (1993); Phys. Lett. B {\bf 324}, 98 (1994).
\bibitem{CL95} C. Corian\`{o} and H-n. Li, Nucl. Phys. {\bf B434}, 535
(1995); C. Corian\`{o}, H-n. Li, and C. Savkli, JHEP {\bf 9807}, 008
(1998).
\bibitem{ZM} M. Tamazouzt, Phys. Lett. B {\bf 211}, 477 (1988);
E. Maina and R. Torasso, Phys. Lett. B {\bf 320}, 337
(1994); D.F. Zeng and B.Q. Ma, Phys. Lett. B {\bf 542}, 55 (2002).
\bibitem{KN} G.R. Farrar and H. Zhang, Phys. Rev. D {\bf 41}, 3349 (1990);
{\bf 42}, 2413 (1990)(E); A.S. Kronfeld and B. Ni\v{z}i\'{c},
Phys. Rev. D {\bf 44}, 3445 (1991).
\bibitem{FSZ} G.R. Farrar, G. Sterman and H. Zhang, Phys. Rev. Lett.
{\bf 62}, 2229 (1989).
\bibitem{LL04} H-n. Li and H.S. Liao, hep-ph/0404050.
\bibitem{Kro} R. Jakob and P. Kroll, Phys. Lett. B {\bf 315}, 463
(1993); B {\bf 319}, 545 (1993)(E).
\bibitem{KLS} Y.Y. Keum, H-n. Li, and A.I. Sanda,
Phys. Lett. B {\bf 504}, 6 (2001); Phys. Rev. D {\bf 63}, 054008 (2001);
Y.Y. Keum and H-n. Li, Phys. Rev. {\bf D63}, 074006 (2001).
\bibitem{PB1} P. Ball, JHEP {\bf 01}, 010 (1999).
\bibitem{BB} P. Ball and M. Boglione, Phys. Rev. D {\bf 68},
094006 (2003).
\bibitem{KR} P. Kroll and M. Raulfs, Phys. Lett. B {\bf 387},
848 (1996).
\bibitem{BMS} A.P. Bakulev, S.V. Mikhailov and N.G. Stefanis, Phys.
Rev. D {\bf 67}, 074012 (2003).
\bibitem{SSK} N.G. Stefanis, W. Schroers, H.-Ch. Kim,
Eur. Phys. J. C {\bf 18}, 137 (2000); A.P. Bakulev, S.V. Mikhailov,
and N.G. Stefanis, Phys. Lett. B {\bf 508}, 279 (2001);
Erratum-ibid. B {\bf 590}, 309 (2004).




\end{thebibliography}
\end{document}